\documentclass[useAMS,usenatbib]{emulateapj}
\usepackage{longtable}

\def\approxgt{\ifmmode \rlap{$>$}{}_{{}_{{}_{\textstyle\sim}}} \else$\rlap{$>$}{}_{{}_{{}_{\textstyle\sim}}}$\fi} 
\def\approxlt{\ifmmode \rlap{$<$}{}_{{}_{{}_{\textstyle\sim}}} \else$\rlap{$<$}{}_{{}_{{}_{\textstyle\sim}}}$\fi}

\def\farcs{\hbox{$.\!\!^{\prime\prime}$}}
\def\degr{\hbox{$^\circ$}}
\def\arcmin{\hbox{$^\prime$}}
\def\arcsec{\hbox{$^{\prime\prime}$}}

\def\flx{erg cm$^{-2}$ s$^{-1}$}
\def\lum{erg s$^{-1}$}

\def\chan{{\it Chandra}}

\shorttitle{The Galactic Bulge Survey}
\shortauthors{Jonker et al.}

\begin{document}

\title{The Galactic Bulge Survey: completion of the X--ray survey observations}

\author{Peter G.~Jonker\altaffilmark{1,2,3}} 
\author{Manuel A.P.~Torres\altaffilmark{1,3}}
\author{Robert I.~Hynes\altaffilmark{4}}
\author{Thomas J.~Maccarone\altaffilmark{5}}
\author{Danny Steeghs\altaffilmark{6}}
\author{Sandra Greiss\altaffilmark{6}}
\author{Christopher T.~Britt\altaffilmark{5,4}}
\author{Jianfeng Wu \altaffilmark{3}}
\author{Christopher B.~Johnson \altaffilmark{4}}
\author{Gijs Nelemans\altaffilmark{2}}
\author{Craig Heinke\altaffilmark{8}}

\altaffiltext{1}{SRON, Netherlands Institute for Space Research, Sorbonnelaan 2,
  3584~CA, Utrecht, The Netherlands}
\altaffiltext{2}{Department of Astrophysics/ IMAPP, Radboud University Nijmegen,
Heyendaalseweg 135, 6525 AJ, Nijmegen, The Netherlands}
\altaffiltext{3}{Harvard--Smithsonian  Center for Astrophysics, 60 Garden Street, Cambridge, MA~02138, USA}
\altaffiltext{4}{Louisiana State University, Department of Physics and Astronomy,
Baton Rouge LA 70803-4001, U.S.A.}
\altaffiltext{5}{Department of Physics, Texas Tech University, Box 41051, Lubbock, TX 79409-1051, USA}
\altaffiltext{6}{Astronomy and Astrophysics, Department of Physics, University of Warwick, Coventry, CV4~7AL, UK}
\altaffiltext{7}{Department of Physics, University of Alberta, Room 238 CEB, Edmonton, AB T6G 2G7, Canada}
\altaffiltext{8}{Institute for Astronomy, KU Leuven, Celestijnenlaan 200D, 3001, Leuven, Belgium}

\begin{abstract} \noindent We provide the \chan\, source list for the
  last $\approx$quarter of the area covered by the Galactic Bulge
  Survey (GBS). The GBS targets two strips of $6\degr\times1\degr$ (12
  square degrees in total), one above ($1\degr<b<2\degr$) and one
  below ($-2\degr<b<-1\degr$) the Galactic plane in the direction of
  the Galactic Center at X--ray, optical and near-infrared
  wavelengths. For the X--ray part of the survey we use 2~ks per
  \chan\, pointing. We find 424 X-ray sources in the 63 \chan\,
  observations we report on here. These sources are in addition to the
  1216 X-ray sources discovered in the first part of the GBS survey
  described before. We discuss the characteristics and the X-ray
  variability of the brightest of the sources as well as the radio
  properties from existing radio surveys. We point out an interesting
  asymmetry in the number of X-ray sources as a function of their
  Galactic $l$ and $b$ coordinates which is probably caused by
  differences in average extinction towards the different parts of the
  GBS survey area.\end{abstract}


\section{Introduction} 

Stellar mass black holes and neutron stars are the end point of
massive star evolution via supernovae or gamma-ray bursts. Nearly all
of the Galactic black holes, and many neutron stars, found so far are
in binaries. Their properties are the observable consequences of
binary interactions. Studying these remnants provides vital clues to
understanding the evolutionary processes that produce them, both in
terms of single massive star evolution, and binary star evolution.
For example, the current stellar mass black hole distribution based on
a sample of about 20 objects appears to be disjoint from that of
neutron stars (\citealt{2010ApJ...725.1918O};
\citealt{2012ApJ...757...55O}; \citealt{2011ApJ...741..103F})
suggesting a bimodality in formation that produces either low--mass
neutron stars or relatively high--mass black holes, with few systems
in between. This remains a challenge for supernova models to reproduce
(\citealt{2012ApJ...749...91F};
\citealt{2012ApJ...757...91B}). \citet{2012ApJ...757...36K} argue that
this mass gap may, in part, be due to systematic effects
underestimating the system inclination.

Unfortunately, our observational sample, particularly in the case of
black holes, is largely comprised of objects discovered in transient
X-ray outbursts, leading to a variety of possible selection effects
that could obscure the properties of the true population (e.g.~see
\citealt{2005ApJ...623.1017N}). For instance, one could envisage,
using the disk instability model including disk irradiation effects
(cf.~\citealt{2008NewAR..51..752L}), an inverse correlation between
the accretor mass and the the duty cycle, reducing the chance of
detection in outburst of relatively low-mass black holes. Additional
selection effects could be invoked by the black hole mass -- orbital
period correlation (\citealt{2002ApJ...575..996L}) and possibly
related to that, the optical and X-ray outburst peak luminosity --
orbital period correlation (\citealt{1998MNRAS.295L...1S} and
\citealt{2010ApJ...718..620W}, respectively).

To mitigate the selection effects incurred by selecting systems that
recently went through an outburst cycle we designed the Galactic Bulge
Survey (GBS; \citealt{2011ApJS..194...18J}). The GBS is a wide,
shallow \chan\, X-ray survey of the Galactic Bulge aiming to uncover
many ($>$100) new {\it quiescent} black hole and neutron star
binaries. As a result, we may find sources quite different to those
identified in outburst. A second goal of the survey is to constrain
binary evolution models (e.g.~\citealt{1999MNRAS.309..253K};
\citealt{2002ApJ...571L..37P}; \citealt{2004ApJ...603..690B}) using
the observed number ratio between $\approx$hundred X-ray binaries and
several hundred CVs that we expect to find. This number will in
particular put constraints on uncertain phases in the binary evolution
such as the common envelope phase (e.g.~;
\citealt{2006MNRAS.369.1152K}; \citealt{2013A&ARv..21...59I}).

For both these science goals we need to classify the X-ray
sources. Given that this classification relies on
multi-wavelength data, by design, the survey area is sufficiently out
of the plane to allow (multi-epoch) optical and near-infrared (NIR)
follow-up of the majority of detected sources. In addition to
classification, optical and NIR spectroscopic observations are also
crucial for dynamical studies to derive compact object masses (and
sometimes the dynamical masses are necessary for classification,
e.g.~\citealt{2013MNRAS.428.3543R}).

The GBS is well under way. Radio counterparts to a sample of sources
from the first part of the X-ray survey have been identified by
\citet{2012MNRAS.426.3057M}. \citet{2012ApJ...761..162H} reported on
associations of X-ray sources with the brightest optical
counterparts. Results from optical variability alone
(\citealt{2012AcA....62..133U}) and optical variability and
spectroscopic studies together (\citealt{2013MNRAS.428.3543R};
\citealt{2013ApJ...769..120B}; \citealt{2013arXiv1310.2597H}; Torres
et al.~2013 submitted) are appearing. Furthermore, we are using NIR
observations from the NIR surveys Two Micron All Sky Survey (2MASS),
VISTA Variables in The Via Lactea (VVV) and the UKIRT Infrared Deep
Sky Survey to identify counterparts to the GBS X-ray sources (Greiss
et al.~2013 submitted).

We here report on \chan\, observations of the final $\approx$quarter
of the sky area of 12 square degrees that makes up the GBS, completing
the \chan\, survey observations of the GBS area. The initial
three-quarters were reported in \citet{2011ApJS..194...18J}. In
addition, we provide the radio counterparts to the X-ray sources
discovered in the final part after \citet{2012MNRAS.426.3057M}
reported on archival radio sources for the first
three-quarters. Finally, we investigate the spatial distribution of
all the X-ray sources found in the GBS area and by comparing with the
ROSAT sources in the sky area we report on here we investigate the
variability properties of the new GBS X-ray sources.

\section{\chan\,X--ray observations, analysis and results} 

\subsection{Source detection}
 
We have obtained 63 observations with the \chan\, X--ray observatory
(\citealt{2002PASP..114....1W})  covering the remaining quarter of
the  total area of twelve square degrees that we call the GBS.
We employed as much as possible the same analysis tools and techniques
as described in Jonker et al.~(2011) in order to come to an as
homogeneous as possible survey. Also we follow the source  naming
convention introduced there, where sources reported in Jonker et
al.~(2011) are referred to as CX\# (after \chan\, X-ray source, where
the numeral indicates the position of that source in the list, with
sources providing the largest number of counts at the detection have
the lowest numeral), while new sources found in the 63 new
observations are called CXB\#.

\begin{figure*}  \hbox{
\includegraphics[angle=0,width=8cm,clip]{ao13-obs-dustmap.ps}
\includegraphics[angle=0,width=12cm,clip]{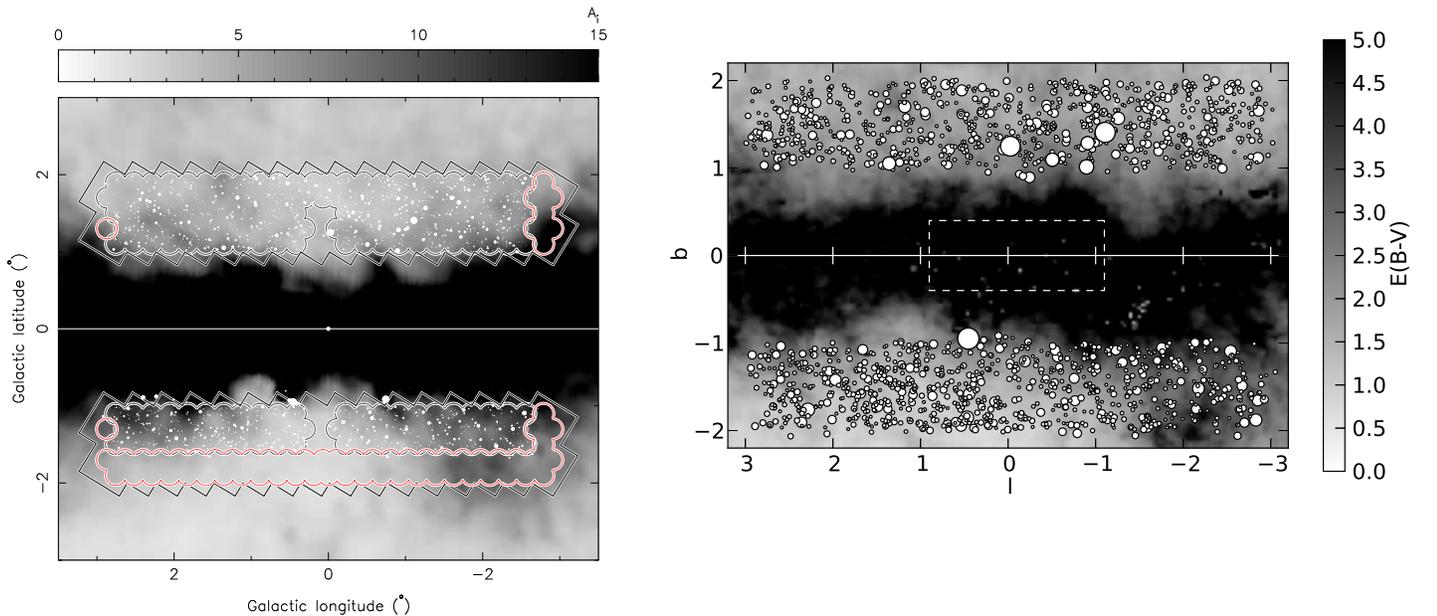}} \caption{{\it
  Left panel:} The large black plus white rimmed saw--tooth boxes are
the outline of our optical observations of the GBS area in Galactic
coordinates. The grey scale image depicts the total reddening in the
Sloan $i^\prime$--band filter, $A_{i^\prime}$, estimated from the
\textsc{Cobe} dust maps \citep{1998ApJ...500..525S}. The overplotted
white circles indicate the position of the \chan\, X--ray sources
detected in the GBS reported in Jonker et al.~(2011). The sources
found in the areas near $l=0^\circ$ and $1^\circ<|b|<2^\circ$ were
reported in Jonker et al.~(2011) but the observations were from
\citet{2009ApJ...706..223H}. The red rimmed curved lines indicate the
composite outline of each circular field of view of 14\arcmin\,
diameter of the 63 \chan\, observations that we report on in this
paper. {\it Right panel:} The grey scale image and contours depict the
total absorption $E(B-V)$, estimated from the extinction maps from the
VVV (\citealt{2012A&A...543A..13G}). The overplotted white circles
indicate the position of all X--ray sources detected in the GBS
including the new sources we report on here. The size of the white
circles is proportional to the number of \chan\, counts detected for
that particular source. The dashed rectangle outlines the region of
the survey of the Galactic Center of
\citet{2002Natur.415..148W}. \\
\\ \\ \\ } \label{changbs} \end{figure*}

In the left panel in Figure~\ref{changbs} we show the 63 new \chan\,
observations we report on here. The red curved line indicates the
composite outline of each circular field of view of 14\arcmin\,
diameter of these 63 observations. The grey curved lines boardering
the white points indicate the composite outline of each circular field
of view of 14\arcmin\, diameter of the individual \chan\, observations
obtained and the detected sources reported in
\citet{2011ApJS..194...18J}, respectively. The area near $l=0^\circ$
is covered by the observations from \citet{2009ApJ...706..223H}.
Sources found in 2 ks-long segments of those exposures were listed in
Jonker et al.~(2011) as well.  In the right panel in
Figure~\ref{changbs} the white circles indicate the position of the
detected point sources. The size of the white circles is an indication
of the number of \chan\, counts detected for that particular source. 

The \chan\,
observations have been performed using the I0 to I3 CCDs of the Advanced CCD
Imaging Spectrometer (ACIS) detector (\citealt{1997AAS...190.3404G};
ACIS--I). The observation identification (ID) numbers for the data presented
here are \dataset [ADS/Sa.CXO#DefSet/GBS] {13528--13590}. We reprocessed and
analyzed the data using the {\sc CIAO 4.3} software developed by the \chan\,
X--ray Center and employing {\sc CALDB} version 4.4.6. The data telemetry
mode was set to {\it very faint} for all observations. The {\it very faint}
mode provides 5$\times$5 pixel information per X--ray event. This allows for
a better screening of events caused by cosmic rays.  In our analysis we
selected events only if their energy falls in the 0.3--8 keV range.

We used {\sc wavdetect} to search for X--ray sources in each of the
observations using data covering the full 0.3--8, the 0.3--2.5 and the
2.5--8 keV energy bands, separately. We set the {\sc sigthresh} in
{\sc wavdetect} to 1$\times 10^{-7}$, which implies that for a
background count rate constant over the ACIS-I CCDs there would be
$<$0.1 spurious source detection per observation as about $1\times
10^6$ pixels are searched per observation. However, in most cases a
source is not detected in a single pixel, thus our estimate of 0.1
spurious source per observation is very conservative. Furthermore, as
we explain below, we applied additional selection criteria.  This
lowers the number of spurious sources further.

We retained all sources for which Poisson statistics indicates that
the probability of obtaining the number of detected source counts by
chance, given the expectation for the local background count rate, is
lower than 1$\times 10^{-6}$. This would be equivalent to a $>5\sigma$
source detection in Gaussian statistics. Next, we deleted all sources
for which {\sc wavdetect} was not able to provide an estimate of the
uncertainty on the right ascension [$\alpha$] and on declination
[$\delta$] as this indicates often that all counts fell in 1 pixel
which could well be due to faint afterglow events caused by cosmic ray
hits. In addition, we impose a 3 count minimum for source detection as
\citet{2005ApJS..161....1M} simulated that in their XBootes survey
with 5 ks ACIS--I exposures, 14 per cent of the 2 count sources were
spurious (note that this percentage will probably be lower for our GBS
exposures of 2 ks).

Since our \chan\, observations were designed to overlap near the
edges, we searched for multiple detections of the same source either
in one of the energy sub-bands or in the full energy band. We consider
sources with positions falling within 5\arcsec\, of each other likely
multiple detections of the same source. This radius is larger than
that of 3\arcsec\, that we took in \citet{2011ApJS..194...18J} as we
found out that still some multiple detections of the same source
remained for sources detected with large off-axis angles (see
\citealt{2012ApJ...761..162H} for the list of 18 sources from
\citealt{2011ApJS..194...18J} that were in fact multiple detections of
the same source.) This means that in \citet{2011ApJS..194...18J} we
found 1216 unique sources. 

In the last quarter of the GBS area that we report on here, we found
that 26 sources are detected more than once. Out of these 26 sources,
23 sources are detected two times, and 3 sources are detected three
times. Two of the sources detected twice were already detected and
reported in \citet{2011ApJS..194...18J} (CX155 and CX314).  We do not
list these two sources in the Table~\ref{srclist} as they were
mentioned in \citet{2011ApJS..194...18J}. The properties that we list
in Table~\ref{srclist} for the sources that are detected multiple
times are those of the detection that gave rise to the largest number
of X--ray counts. In Table~\ref{srclist} we also list the number of
times that sources are detected.

Besides the multiple detections of CX155 and CX314 fourteen additional
sources detected once in the Cycle 13 \chan\, observations were previously
detected and listed in \citet{2011ApJS..194...18J}. These sources are CX15,
CX17, CX25, CX44, CX60, CX69, CX79, CX137, CX221, CX266, CX312, CX355, CX374,
CX439. In most cases the off-axis angle of the source position was larger
during the new observations and, given that a similar number of X--ray counts
was detected in each instance, the source position provided in
\citet{2011ApJS..194...18J} is the most accurate X--ray position available.
The main exception where we consider the newly derived position to be more
accurate is CX314. CX314 was detected at 10.8\arcmin\, off-axis at 8 counts
in the \chan\, detection leading to its discovery. The new detection we
report on here provides 17 counts and the source was 5.9\arcmin\, off-axis in
ObsID 13581. The new best-fit source position is ($\alpha$,
$\delta$)=(266.6461515,-31.8136964) which is 2.6\arcsec\, from the
previously reported position.

Others, like CX25, were detected closer on axis in the new cycle 13
observations (6.7\arcmin\, off--axis with 6 counts) but with much more
counts in the observation reported in \citet{2011ApJS..194...18J}
(7.2\arcmin\, off-axis with 48 counts) than in the new cycle 13
observation implying that the position provided in
\citet{2011ApJS..194...18J} will be more accurate. We do conclude that
CX25 is variable in X-rays.

In total we detected 424 distinct sources in the area indicated with
red circles and the red curved lines on the left side in
Figure~\ref{changbs}. The source list is given in Table~\ref{srclist}
and the table provides information on $\alpha$, $\delta$, the error on
$\alpha$ and $\delta$, total number of counts detected, the
observation ID of the observation resulting in the detection and the
off-axis angle at which the source is detected. The error on $\alpha$
and $\delta$ are the error provided by {\sc wavdetect}, it does not
take  into account the typical \chan\, bore--sight uncertainty of
0.6\arcsec\, (90 per cent confidence). We do, however, add a column to
Table~\ref{srclist} quoting the total uncertainty on the source
position following formula 4 in \citet{2010ApJS..189...37E}. For
clarity, we repeat their equation here below, \begin{displaymath} \log
P = \left\{ \begin{array}{llr} 0.1145 \theta -  0.4957 &\log C   +
0.1932 & \\ & \mathrm{for\,} 0.0 <\log C < 2.1393 & \\ 0.0968\theta - 
0.2064 &\log C   - 0.4260 & \\ & \mathrm{for\, } 2.1393<  \log C < 3.3
& \\ \end{array} \right. \end{displaymath} where $\theta$ is the
off-axis angle in arcminutes and $C$ is the detected number of X-ray
photons. The positional error $P$ is given in arcseconds and it
corresponds to a 95\% confidence interval.

We provide individual \chan\, source names, however, for briefness we use
the source number in Table~\ref{srclist} preceded by "CXB" to indicate
which source we discuss in this paper. For the error $\sigma_N$ on the
detected number of counts $N$, \citet{2005ApJS..161..271G} give $\sigma_N
= 1+\sqrt{N+0.75}$ after \citet{1986ApJ...303..336G}. To allow for an
rough, easy calculation of the source flux based on the detected number of
source counts we give the conversion factor for a source spectrum of a
power law with photon index of 2 absorbed by $N_{\rm H}=1\times 10^{22}$
cm$^{-2}$: $7.76\times 10^{-15}$~\flx\,photon$^{-1}$.

\renewcommand{\arraystretch}{2.0}
\begin{center}
\begin{longtable*}{cccccccccccc}

  \caption{PLACEHOLDER, FIRST TEN ENTRIES ONLY! The GBS X--ray source
    list providing the GBS source name, the source number as used in
    this paper is preceded by "CXB" to differentiate it from the
    sources in Jonker et al.~(2011), $\alpha$, $\delta$ in decimal
    degrees, the $3\sigma$ error on localizing the source on the detector
    $\alpha$ and $\delta$ in arcseconds, total number of counts
    detected, the ID of the observation resulting in the
    detection, the off-axis angle at which the source is detected, the
    number of times the source was detected in the \chan\,
    observations, the 95\% confidence positional uncertainty
    ($\Delta$pos) calculated according to formula 4 in
    \citet{2010ApJS..189...37E} taking the boresight uncertainty into
    account, and the hardness ratio (HR) for sources detected with
    more than 20 counts. The hardness is defined as the ratio between
    the count rate in the 2.5--8 keV minus that in the 0.3--2.5 keV
    band to the count rate in the full 0.3--8 keV energy band. The HR
    is calculated for the detection where the off-axis angle was
    smallest if the source was detected multiple times.}

\label{srclist}\\
\hline
Source & CXB\# & $\alpha$& $\delta$ & $\Delta \alpha$ &
$\Delta\delta$ & \#  & Obs & Off-axis & \# of & $\Delta$pos & HR \\[0.1mm]
name &  & (degrees) & (degrees) & (\arcsec) & (\arcsec) & (cnt) & ID &
angle (\arcmin) & detec. & (\arcsec) & \\[0.1mm]

CXOGBSJ175748.7-275214 & CXB1 & 269.4529160 & -27.8707194 & 0.19 & 0.22 & 161 & 13536 & 7.74 & 1 & 0.74 & -0.61$\pm$0.06  \\ 
CXOGBSJ175359.8-292907 & CXB2 & 268.4994759 & -29.4852781 & 0.09 & 0.05 & 148 & 13550 & 4.35 & 2 & 0.35 & -0.18$\pm$0.02  \\ 
CXOGBSJ174614.3-321949 & CXB3 & 266.5599883 & -32.3303786 & 0.06 & 0.05 & 105 & 13574 & 2.64 & 1 & 0.31 & 0.28$\pm$0.03  \\ 
CXOGBSJ173416.2-304538 & CXB4 & 263.5678548 & -30.7607505 & 0.15 & 0.09 & 70 & 13586 & 3.78 & 1 & 0.51 & -0.90$\pm$0.12  \\ 
CXOGBSJ173208.6-302828 & CXB5 & 263.0362304 & -30.4746348 & 0.07 & 0.10 & 66 & 13587 & 3.78 & 1 & 0.53 & -0.75$\pm$0.10  \\ 
CXOGBSJ174517.0-321356 & CXB6 & 266.3208565 & -32.2323620 & 0.11 & 0.11 & 66 & 13577 & 3.73 & 2 & 0.52 & 0.78$\pm$0.11  \\ 
CXOGBSJ175551.6-283213 & CXB7 & 268.9650346 & -28.5369772 & 0.06 & 0.05 & 65 & 13533 & 1.83 & 1 & 0.32 & 0.34$\pm$0.05  \\ 
CXOGBSJ175432.1-292824 & CXB8 & 268.6339299 & -29.4734138 & 0.28 & 0.26 & 65 & 13550 & 7.49 & 2 & 1.42 & -0.78$\pm$0.11  \\ 
CXOGBSJ174916.6-311518 & CXB9 & 267.3192034 & -31.2550666 & 0.09 & 0.07 & 64 & 13569 & 3.52 & 1 & 0.50 & -0.95$\pm$0.13  \\ 
CXOGBSJ175832.4-275244 & CXB10 & 269.6350093 & -27.8789043 & 0.13 & 0.11 & 53 & 13558 & 4.30 & 1 & 0.68 & -0.56$\pm$0.09  \\

\end{longtable*}
\end{center}
\renewcommand{\arraystretch}{1.0}

\subsection{X--ray spectral information}

\begin{figure} \includegraphics[angle=0,width=7cm]{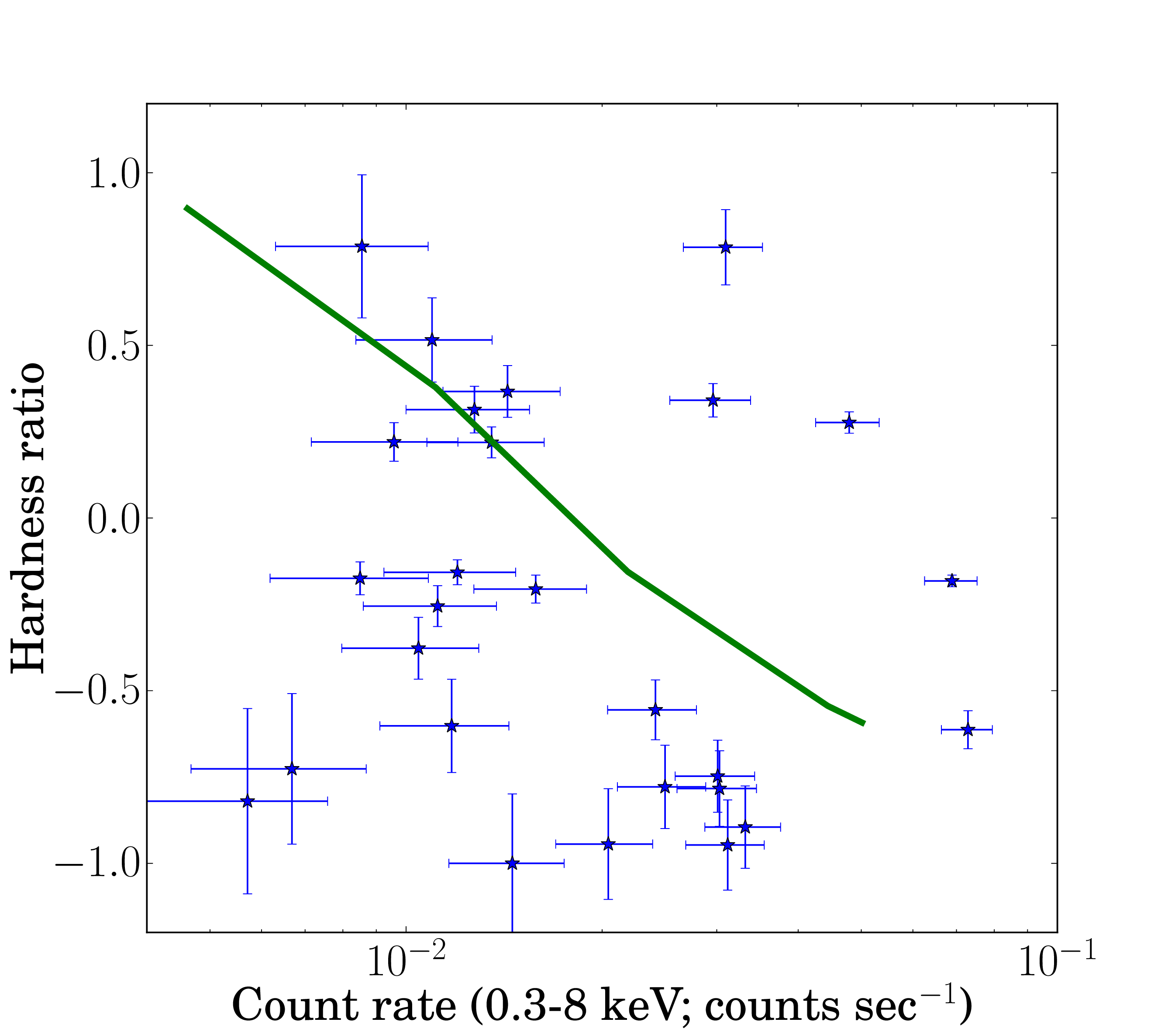}
  \caption{The hardness -- intensity diagram for the 27 sources for
    which 20 or more counts were detected in \chan\, cycle 13
    observations for the GBS survey. To mitigate effects of small
    differences in exposure times we used count rates as a measure of
    intensity. The hardness is defined as the ratio between the count
    rate in the 2.5--8 keV minus that in the 0.3--2.5 keV band to the
    count rate in the full 0.3--8 keV energy band. Hard sources fall
    in the top half and soft sources in the bottom half of this
    figure. The green line shows the influence of the
    extinction (N${\rm_H}$) on a power law spectrum with index 2 for a
    source count rate of 0.05 counts s$^{-1}$ and N${\rm_H}$ values
    increasing from bottom right to top left from (0.01, 0.1, 1, 3,
    10)$\times 10^{22}$ cm$^{-2}$. } \label{hardnessinten} \end{figure}

We extract source counts using circular source extraction regions of
10\arcsec.  Background extraction regions are annulli with inner and outer
radii of 15\arcsec\, and 30\arcsec, respectively. We plot the 27 sources
for which we detected 20 or more counts in a hardness -- intensity diagram
(Figure~\ref{hardnessinten}). To mitigate the effects that small
differences in exposure time across our survey can have, we use count rates
as a measure of intensity. We define the hardness ratio as the ratio
between the count rate in the 2.5--8 keV minus that in the 0.3--2.5 keV
band to the count rate in the full 0.3--8 keV energy band
(after~\citealt{2004ApJS..150...19K}).  We derived the hardness using {\sl
XSPEC} version 12.7 (\citealt{1996adass...5...17A}) by determining the
count rates in the soft and hard band taking the response and ancillary
response file for each of the sources.  For these 27 sources photon
pile--up is less than 10\% even for the brightest source. Naively, one
would expect most hard sources to be more distant and more absorbed than
the soft sources, as the intrinsic spectral shape of the most numerous
classes of sources we expect to find does not differ much.

The most interesting aspect from Figure~\ref{hardnessinten} is perhaps the
presence of three bright (rate $>$2.5$\times 10^{-2}$ counts s$^{-1}$)  and
relatively hard sources (HR$>$0). Their relatively hard spectrum makes it 
likely that these three sources (CXB3 [HR=0.28$\pm$0.03],  CXB6
[HR=0.78$\pm$0.11], and CXB7 [HR=0.34$\pm$0.05]) suffered significantly 
from X-ray absorption thus they likely are at a distance  of more than 3
kpc which given their relatively high X-ray flux means that their  X-ray
luminosity is substantial. CXB3 is probably a transient source (see below)
and none of the three sources is associated with archival radio emission
(see below) decreasing the chance  that they are background AGN, and making
them potential X-ray binaries.

As foreseen, the spectral information is insufficient for source
classification for the majority of the total number of detected
sources, therefore, classification will have to come from
(multi-epoch) multi-wavelength observations.  Finally, there seems to
be a dichotomy in the hardness with one peak centered on a hardness of
0.2 and another centered on -0.8 with a paucity of sources with
hardness 0. A similar dichotomy was reported in
\citet{2011MNRAS.413..595W} and \citet{2011ApJS..194...18J} (see the
latter paper for a possible explanation for the nature of this
dichotomy).

\subsection{\chan\,light curves source CXB\#1--10}

We inspect the \chan\, light curves of source CXB\#1-10. We rebinned
the light curves in 200~s bins. Source CXB\#1, 2, 3, 6 and 9 show
suggestive evidence for flare-like variability.  Fitting the light
curve with a constant gives a $\chi^2$ value of 16 (for 10 degrees of
freedom [d.o.f.]), 35.9 (9 d.o.f.), 19.5 (10 d.o.f.), 18 (10 d.o.f.),
16.4 (9 d.o.f.), respectively. The light curves of source CXB\# 4, 5,
7, 8 and 10 are consistent with being constant with $\chi^2$ values of
8.4 (10 d.o.f.), 7.5 (9 d.o.f.), 11 (10 d.o.f.), 10 (9 d.o.f.) and 3.8
(10 d.o.f.), respectively. We do note that the number of counts in
each 200~s bin varies between 35 and 3 counts between these sources
and as a function of time. Therefore, certainly for the bins
containing only a few counts the use of the $\chi^2$ statistic is
suspect. The small number of counts per bin in several cases makes it
likely that some of the high values of reduced $\chi^2$ are occuring
due to chance fluctuations.

\begin{figure} \includegraphics[angle=0,width=9cm]{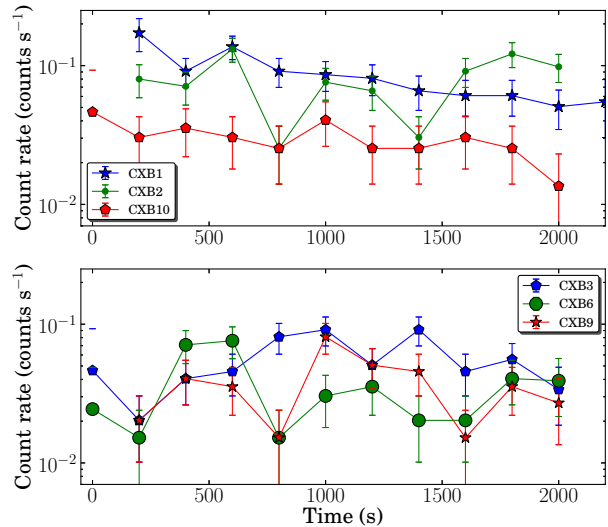}
  \caption{The two panels show the \chan\, X-ray light curves of six
    CXB sources. Each point is an average of 200~s of \chan\,
    data. For five sources there is suggesive evidence that the source
    is variable during the \chan\, observation (CXB1, CXB2, CXB3,
    CXB6, and CXB9). For comparison we plot in the top panel also the
    light curve of CXB10 for which we find no evidence that the source
    varied during the observation.} \label{lightcurves} \end{figure}

In Figure~\ref{lightcurves} we plot the light curves of the sources
for which there is evidence for variability during the
observations (i.e.~CXB1, CXB2, CXB3, CXB6, and CXB9) and for
comparison we plot in the top panel of the same Figure the light curve
of CXB10 for which our current data provides no evidence that the
source varies during the observation.

\section{Discussion}

Using 63 \chan\, observations we cover the remaining $\approx$quarter
of the 12 square degrees that comprise the Galactic Bulge Survey
(\citealt{2011ApJS..194...18J}). In this paper we provide the list of
424 X-ray sources that we find in this area and that have three or more 
counts in the short (2 ks) \chan\, observations.

\begin{figure} \includegraphics[angle=0,width=9cm]{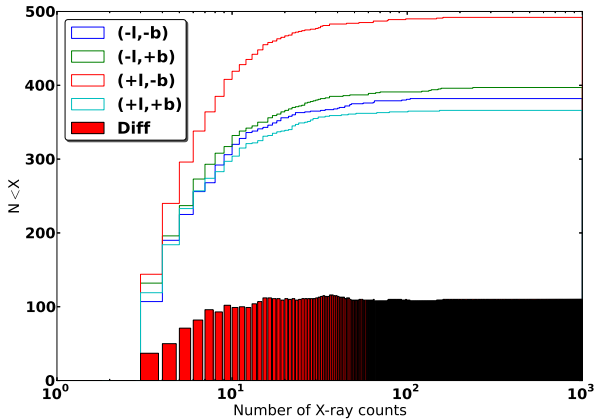}
  \caption{The cumulative distribution of the number of X-ray sources
    as a function of the number of source X-ray counts discovered in
    the GBS for four different quadrants according to the Galactic
    coordinates of the sources (-l,-b), (-l,+b), (+l,+b) and
    (+l,-b). The full histogram shows the cumulative difference in the
    number of X-ray sources as a function of the number of detected
    source X-rays found in the (+l,-b) and the (-l,-b) quadrant. The
    difference is qualitatively the same when comparing the number of
    X-ray sources in the (+l, -b) and the other quadrants. There is a
    clear excess of the number of X-ray sources discovered in the
    (+l,-b) quadrant when compared with the other quadrants. The
    difference increases with X-ray count rate up to sources with
    $\approxlt$10 X-ray counts per source.
  } \label{distri} \end{figure}

In total we detected 1640 unique X-ray sources. Of these 875 are
detected at Galactic latitudes below the plane and 765 at Galactic
latitudes above the plane. For a symmetric distribution of 1640
sources one would expect 820$\pm$20 on either side, making the
detected distribution marginally skewed. We investigated the nature of
this asymmetry by dividing the number sources over the four quadrants
they were detected in. We made quadrants according to the Galactic
coordinates of the source and we counted the number of sources in each
quadrant (-l,-b: \#382), (-l,+b: \#399), (+l,+b: \#366) and (+l,-b:
\#493). It turns out that the quadrant (+l,-b) is responsible for the
apparent asymmetry in the number of detected sourcs (see Figure~\ref{distri}).

Most of the sources we expect to have detected are relatively nearby
(within 3 kpc; \citealt{2011ApJS..194...18J}), nevertheless, the
different average extinction in the GBS areas in the four quadrants
could still have a significant influence on the number of detected
sources. The average extinction is indeed lower in the (+l,-b)
quadrant where we detected most new X-ray sources (cf.~the right panel
of Figure\ref{changbs}). The overdensity of sources we find in
quadrant (+l,-b) of the GBS area coincides with the presence of
diffuse X-ray emitting gas in that part of the GBS area, as found by
ROSAT (\citealt{1997ApJ...485..125S}).

\begin{figure} \includegraphics[angle=0,width=9cm]{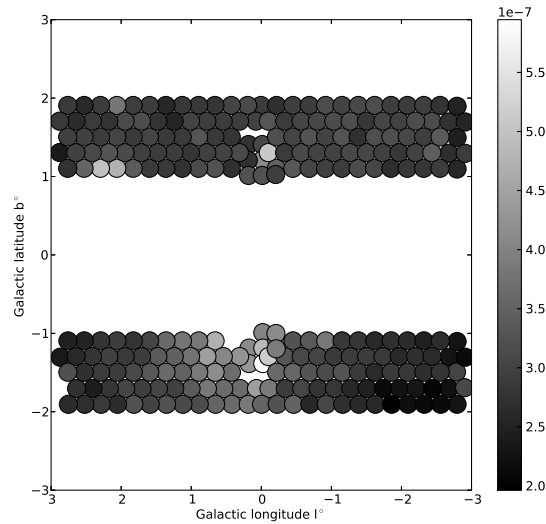}
  \caption{The background count rate (pixel$^{-1}$ s$^{-1}$) as
    measured by \chan. The background is higher in the (+l,-b) part of
    the GBS area than in the other areas. We removed two observations
    from this plot; one where we used the FAINT event mode that does
    not allow for the thorough cleaning of cosmic ray afterglow events
    and therefore yields a much higher background and one where the
    background is artificually increased due to the presence of a very
    bright X-ray source.  } \label{backchan} \end{figure}

In order to investigate this asymmetry further we compared the
different background levels in our \chan\ observations as determined
by the {\sc wavdetect} tool (see Figure~\ref{backchan}, a higher
background is indicated by a lighter shade of gray). The background
levels could influence the detection probability especially for
sources with 3 counts falling far away from the optical axis of the
satellite. The diffuse emission could show up as a diffuse number of
pixels with 1 or 2 counts or in areas with a lower extinction a larger
amount of 1 and 2 cnt sources such as RS CVn and coronally active
stars can be present.

For a background count rate per pixel per second of $\approx5\times
10^{-7}$ (see Figure~\ref{backchan}) and 2 ks.~exposures and
$\approxlt$100 pixels for the point spread function far off axis, the
expected background rate is $\approxlt$0.1 count per 2 ks.~observation
in such an area. Whereas there is indeed a difference in the
background count rate in line with the expectation from either more
1--2 count point sources or more diffuse emission in the (+l,-b)
quadrant of the GBS area, this enhanced background does not have a
large effect on the number of 3 count sources even far off-axis.

We conclude that the overdensity of sources in the (-l,+b) part of our
GBS area is likely caused by the lower average extinction in that
quadrant of the GBS survey area, whereas the higher X-ray background
in that area is in line with the diffuse gas as found by
\citet{1997ApJ...485..125S}. Those authors argued that this diffuse
gas is at the distance of the bulge.

\subsection{Comparison with ROSAT sources}

In order to investigate whether sources in our CXB source list are detected by
ROSAT we cross-correlated the GBS CXB source list with the ROSAT All Sky
Survey (RASS; \citealt{1999A&A...349..389V}). We queried both the Bright as
well as the Faint catalog, and the ROSAT High Resolution Imager (HRI)
Pointed Observations (1RXH), and the Second ROSAT Position Sensitive
Proportional Counter (PSPC) Catalog (2RXP) using the VizieR database. To
accommodate the relatively large positional uncertainties in many of the
ROSAT source detections, we searched for ROSAT sources within 30\arcsec\,of
the \chan\, positions of our CXB sources.

We find two RASS (faint) sources that have a position relatively close to
the GBS CXB sources CXB9 and CXB11. These sources are probably associated
with the ROSAT sources 1RXS~J174916.5-311509 and 1RXS~J175019.0-302654,
respectively.  CXB9 is 9.3\arcsec\, away from 1RXS~J174916.5-311509. CXB9
is also associated with an O8~III Tycho-2 source
(\citealt{2012ApJ...761..162H}, see their work for further details on this
source). CXB11 is 22\arcsec\, away from 1RXS~J175019.0-302654 which is
probably the same source as 2RXP~J175020.0-302616. We furthermore find that
CXB55 is likely to correspond to 1RXH~J175017.6-311427 (reported in
\citealt{1994ApJ...423..633R}). The angular distance between the two
sources is 12.2\arcsec. Finally, CXB93 might be related to 1RXH
J174612.7-320637 which is located at an angular separation of 25\arcsec.

\subsection{Transient sources}

The first three CXB sources (CXB1--3) are bright enough that they should
have been detected in the RASS if they were as bright during the RASS as
they are in our \chan\, observations. However, they were not detected in
the RASS, and thus we are inclined to conclude that their X-ray luminosity
has significantly varied between our \chan\, and the RASS observation.
Before we firmly conclude that these sources are variable, we
verified the \chan\, X-ray spectrum of each of these sources. CXB1 and CXB2
have spectra that should have allowed for a detection in RASS, however, we
found that the spectrum of CXB3 is strongly absorbed potentially providing
an explanation as to why ROSAT did not detect the source. Using
C-statistics we fit a spectral model consisting of a power-law absorbed by
interstellar material to the X-ray spectrum. For CXB3 we find a best fit
N$_H=(2.7\pm 0.9)\times 10^{22}$ cm$^{-2}$ for a power law index of $2.4\pm
0.7$. Extrapolating this model to the ROSAT band (0.01-2.5 keV) we find
that the source flux is 2.5$\times 10^{-12}$\flx. This implies that the
source should have been detected by the RASS although we note that the
extrapolation to low energies carries a significant uncertainty. We
tentatively conclude that CXB1,  2, and 3 are transient or at least
highly variable sources.

CXB3 has a bright near-infrared counterpart of $K=10.06\pm0.04$
(2MASS~J17461440$-$3219494; this 2MASS source was not picked-up in our
cross-correlation with {\sc simbad} see Section~\ref{optical}) at an angular
distance of 0\farcs13, which is consistent with the 95\% confidence
uncertainty on the position of the source of 0\farcs31 (see
Table~\ref{srclist}). The extinction towards the source as given by
\citet{2012A&A...543A..13G} is $E(B-V)\sim 2.8$. This yields an
N$_H\sim1.6\times 10^{22}$ cm$^{-2}$ which is consistent within the
uncertainties with the value we find from our fit to the X-ray
spectrum (using the conversion of $E(B-V)$ to $A_V$ using a gas to
dust ratio of $R=3.1$ and the conversion from $A_V$ to N$_H$ from
\citealt{1995A&A...293..889P}). This value for the extinction is also
consistent with a distance to the source of $\sim 8$~kpc. For that
distance the source luminosity will be around $6\times10^{33}$\lum. The source
is also detected in the 2MASS, WISE and GLIMPSE surveys
(\citealt{2006AJ....131.1163S}; \citealt{2010AJ....140.1868W};
\citealt{2003PASP..115..953B} and \citealt{2009PASP..121..213C},
respectively) as well as in our Blanco/DECam $r^\prime$ data at
$r^\prime\sim19.8$ (Johnson et al. in prep). Correcting for the
reddening of \citet{2012A&A...543A..13G} we find that the spectral
energy distribution fits well with the Kurucz model
(\citealt{2004astro.ph..5087C}) of late K red giant of T$_{eff}=4000$K
and $\log g=1.5$. This source is a candidate symbiotic X-ray binary
(cf.~\citealt{2013arXiv1310.2597H}).

We also investigated whether ROSAT sources found using the RASS (Bright and
Faint catalogues) as well as pointed observations (from the HRI and the
PSPC) fall in the observed GBS CXB area but were not detected. We use {\sc
Topcat} to cross-correlate the VizieR ROSAT catalogues mentioned above with
the coordinates of the \chan\, pointing centers. We considered a sky area
of 7\arcmin\, around the \chan\, pointing centers in this
cross-correlation. The resulting list contains all the ROSAT sources that
fall inside this sky area. We remove the ROSAT sources that have an
associated GBS CXB counterpart within 30\arcsec\,(see above). Below we
discuss the ROSAT sources that were no longer detected in the GBS CXB
observations.

1RXH~J174423.1-320254 and 1RXH~J174449.9-321701 have no CXB
counterpart within 30\arcsec, however, both sources were detected by
ROSAT at a signal-to-noise ratio of only 3 and 2.7, respectively.  The
Second ROSAT PSPC Catalog source 2RXP~J175138.6-295024 also went
undetected in the CXB. The false alarm probability for the ROSAT
detection of this source is 1.2$\times 10^{-2}$.

There are 15 sources from the RASS Faint source catalog within 7\arcmin\,
of a \chan\, CXB pointing that do not have a CXB counterpart within
30\arcsec\, (see Table~\ref{faintrass}). However, we note that the
uncertainty on the position of the faint RASS sources ranges between
14-49\arcsec\,and the search radius of 30\arcsec\, might be too strict.
However, enlarging the matching radius provides other problems. E.g.~for a
search radius of 1\arcmin, 1RXS~J174608.8-320544 has two potential CXB
counterparts CXB93 and CXB406. CXB93 is at 57.2\arcsec\, and CXB406 lies at
51.6\arcsec\, from 1RXS~J174608.8-320544 (CXB93 and CXB406 are
94.7\arcsec\, apart and they are thus not consistent with being the same
source). Interestingly, given that 1RXS~J174608.8-320544 is marked as a
potentially  extended source in the RASS catalog, it might be that the
source is a blend of CXB93  and CXB406. 

\begin{table} \caption{RASS faint sources without CXB X-ray counterparts
within 30\arcsec. $\Delta$WCS is the uncertainty on the source position
provided by the RASS. $L$ is the likelihood of source detection $L =
-\ln(1-P)$, where $P$ is the probability that the source is real. Those
sources with $L\approxgt 9$ that went  undetected in the GBS are good
candidate transients. }

\label{faintrass}
\begin{center}
\begin{tabular}{lcccc}
1RXS   &  RA &DEC & $\Delta$WCS (\arcsec)& $L$\\
J175237.6-294714$^a$ &268.1567 &-29.78722 &49& 9 \\
J175343.3-291444 &268.4304 &-29.2457  &16& 8 \\
J175342.4-290809 &268.4267 &-29.1358 &19& 10\\
J175420.8-285412 &268.5867 &-28.9033 &15& 12\\
J175606.4-283311 &269.0267 &-28.5532 &30&  8\\
J175712.8-280510 &269.3033 &-28.0863 &17& 10\\
J175836.1-273358 &269.6504 &-27.5661 &19&  8\\
J175050.7-301735 &267.7112 &-30.2932 &37&  7\\
J175323.2-295649 &268.3467 &-29.9471 &19&  8\\
J175334.9-295013 &268.3954 &-29.8369 &14&  8\\
J175421.9-292206 &268.5913 &-29.3683 &14& 15\\
J175855.9-272945 &269.7329 &-27.4960 &27& 9 \\
J175019.0-304843 &267.5792 &-30.8119 &30& 10\\
J174906.7-311915 &267.2779 &-31.3208 &21& 11\\
J174608.8-320544$^a$ &266.5367 &-32.0956 &25& 17\\
\end{tabular}
\end{center}
{\footnotesize $^a$ Marked in the RASS as a potentially extended ROSAT source.}
\end{table}

For all the sources with $L\approxgt 9$ in Table~\ref{faintrass} as
well as the two 1RXH sources and the one 2RXP source not detected in
CXB, it is conceivable that the ROSAT observations found the source in
a bright state and/or that the source spectrum is too soft to allow
for a detection in the GBS CXB observations.  Several sources present
secure ROSAT detections and they should have been detected in our CXB
observations. E.g.~1RXS~J175421.9-292206 is detected at more than
5$\sigma$ significance with ROSAT hardness ratio 1 (HR1)=$0.43\pm
0.37$ and hardness ratio 2 (HR2)=$0.22\pm0.42$. Here, HR1= (B-A)/(B+A)
and HR2= (D-C)/(D+C), with A=0.11-0.41 keV, B=0.52-2.0 keV, C=0.5-0.9
keV, and D=0.9-2.0 keV count rate. Therefore, the X-ray spectrum is
not too soft for \chan, indicating that this source has varied between
the ROSAT and the \chan\, observations.  For some other sources, most
notably those with L$\approxlt9$ in Table~\ref{faintrass}, the ROSAT
detection significance is also so low that they could be spurious
detections.

\section{Radio NVSS detections and {\sc simbad} listing of GBS CXB sources}
\label{optical}

After \citet{2012MNRAS.426.3057M} we provide the result from the
cross-correlation between the CXB source list and the NRAO VLA Sky Survey
(NVSS), where NRAO and VLA stand for National Radio Astronomy Observatory
and Very Large Array, respectively. We considered sources within
30\arcsec\, of a CXB source as a likely match. Table~\ref{radio} contains
the nine NVSS sources we find and their likely CXB counterpart.

The three radio bright objects associated with CXB23, CXB127 and CXB150 are
also detected in \citet{2004AJ....128.1646N} as 330 MHz sources called
GCPS~359.845-1.845 ($\Delta=3.8$\arcsec; S$_{330 {\rm MHz}}=764$mJy),
GCPS~358.154-1.680 ($\Delta=17$\arcsec; S$_{330 {\rm MHz}}=1464$mJy), and
GCPS~359.912-1.815 ($\Delta=3.6$\arcsec; S$_{330 {\rm MHz}}=474$mJy),
respectively. For $S_\nu \propto \nu^\alpha$, where $\nu$ is the radio
frequency $S_\nu$ the radio flux this yields $\alpha=-0.5,-0.7,-0.5$,
respectively. These sources have radio spectra consistent with being Active
Galactic Nuclei and we thus preliminary classify CXB23, CXB127 and CXB150
as such.

For the other GBS CXB sources with potential radio counterparts it is more
difficult to provide a classification on the basis of the potential
association with the radio source alone.

\begin{table*}
\caption{NVSS sources close to CXB X-ray sources. }
\begin{center}
\label{radio}
\begin{tabular}{lccccccc}
  \multicolumn{1}{l}{CXB\#} &
  \multicolumn{1}{c}{NVSS} &
  \multicolumn{1}{c}{RA (J2000)} &
  \multicolumn{1}{c}{Dec (J2000)} &
  \multicolumn{1}{c}{$\Delta$RA (sec)} &
  \multicolumn{1}{c}{$\Delta$DEC (\arcsec)} &
  \multicolumn{1}{c}{S1.4 (mJy)} &
  \multicolumn{1}{c}{Separation (\arcsec)} \\
\hline
CXB19& 175737-281000 & 17 57 37.72 & -28 10 00.7 & 0.38 & 9.4 & 3.8$\pm$ 0.6 & 7.9\\
CXB23&175230-300107 & 17 52 30.97 & -30 01 07.8 & 0.03 & 0.6 & 350$\pm$10 & 0.95\\
CXB28&175205-303026 & 17 52 05.68 & -30 30 26.7 & 0.43 & 6.7 & 2.8$\pm$0.5 & 3.6\\
CXB127&174748-312315 & 17 47 48.62 & -31 23 15.2 & 0.04 & 0.6 & 540$\pm$ 15 &  5.0\\
CXB150&175233-295645 & 17 52 33.16 & -29 56 45.5 & 0.03 & 0.6 & 235$\pm$10 &  0.71\\
CXB162&173357-302729 & 17 33 57.85 & -30 27 29.2 & 0.14 & 1.9 & 8.3$\pm$0.5 & 1.37\\
CXB163&173229-302522 & 17 32 29.34 & -30 25 22.7 & 0.62 & 8.6 & 2.4$\pm$0.6 & 17.9\\
CXB288&173251-302919 & 17 32 51.74 & -30 29 19.4 & 0.05 & 0.7 & 35$\pm$1.2 & 1.1\\
CXB384&174857-310445 & 17 48 57.10 & -31 04 45.4 & 0.46 & 8.3 & 5.5$\pm$0.7 & 3.5\\
\end{tabular}
\end{center}
\end{table*}

\begin{table*} 
  \caption{Optical or near-infrared sources found in the {\sc simbad} data base within 5\arcsec\,of CXB X-ray sources. Radio or
    X-ray sources found in the {\sc simbad} data base within 30\arcsec\,of CXB
    X-ray sources. Angular D stands for the angular distance between the
    {\sc simbad} and the CXB source position. PM* means high proper motion
    star, EB* stands for eclipsing binary star. V* denotes variable
    star and ** means double or multiple star. PN stands for planetary
    nebula and YSO for young stellar object. Finally, supernova
    remnant is abreviated by SNR and Cataclysmic Variable by CV.}

\begin{center}
\label{simbad}
\begin{tabular}{llccccccc}
  \multicolumn{1}{l}{CXB\#} &
  \multicolumn{1}{l}{CXB\#} &
  \multicolumn{1}{c}{RAJ2000-CXB} &
  \multicolumn{1}{c}{DECJ2000-CXB} &
  \multicolumn{1}{c}{Angular D} &
  \multicolumn{1}{c}{Simbad name} &
  \multicolumn{1}{c}{RAJ2000} &
  \multicolumn{1}{c}{DEJ2000} &
  \multicolumn{1}{c}{ID} \\
\hline
1 & CXB2  & 268.499460 &-29.4852781 &7.2  &  AX J1754.0-2929             &268.500000 &-29.483333 &X     \\              
2 & CXB5$^a$  & 263.0362304 &-30.474635  &0.2  &  HD 315961                   &263.036154 &-30.474636 & K5   \\             
3a & CXB9$^a$  & 267.3192035 &-31.2550666 &0.6  &  HD 161853                   &267.319017 &-31.255022 & O8~III     \\           
3b & CXB9  & 267.3192035 &-31.2550666 &3.4  &  PN RPZM   40                &267.319583 &-31.254167 &PN?        \\         
3c & CXB9  & 267.3192035 &-31.2550666 &10 &  1RXS J174916.5-311509       &267.318671 &-31.252244 &X           \\        
4 & CXB10 & 269.6350093 &-27.8789043 &0.7  &  MACHO 401.48296.2600        &269.635208 &-27.879000 &CV   \\        
5 & CXB11 & 267.5862652 &-30.4477944 &22 &  1RXS J175019.0-302654       &267.579158 &-30.448469 &X            \\       
6 & CXB17$^a$ & 268.6255656 &-29.3992464 &0.2  &  2MASS J17543011-2923572     &268.625488 &-29.399244 &IR        \\      
7 & CXB21 & 268.7011304 &-29.3277772 &3.7  &  OGLE BUL-SC4 568004         &268.700458 &-29.328611 &V*            \\      
8a& CXB23 & 268.1288255 &-30.0186408 &0.6  &  [IBR2011] J1752-3001        &268.128960 &-30.018515 &Radio         \\      
8b& CXB23 & 268.1288255 &-30.0186408 &1.5  &  [LKL2000] 43                &268.129167 &-30.018333 &Radio \\              
8c& CXB23 & 268.1288255 &-30.0186408 &4.3  &  GCPS  111                   &268.130208 &-30.018500 &Radio   \\            
9& CXB26 & 268.4491784 &-29.7439772 &1.0  &  OGLE BUL-SC3   6033         &268.448875 &-29.743861 &CV  \\         
10& CXB28 & 268.0240465 &-30.5064844 &2.1  &  2XMM J175205.6-303023       &268.023375 &-30.506556 &X         \\          
11& CXB29 & 268.5549195 &-29.4830887 &0.6  &  OGLE BUL-SC4 155897         &268.554750 &-29.483028 &V*        \\          
12 & CXB34 & 266.8706341 &-32.2448156 &12 &  2MASS J17472806-3214462     &266.866917 &-32.246194 &X          \\         
13& CXB36$^a$ & 266.5600100 &-32.1033654 &4.1  &  LTT  7073                   &266.560160 &-32.102233 &PM* M2~V        \\         
14& CXB49 & 267.3703237 &-31.3067944 &0.8  &  2MASS J17492885-3118237     &267.370225 &-31.306603 &Candidate YSO \\      
15& CXB54 & 268.1172224 &-29.9895816 &13 &  RRF    9                    &268.114167 &-29.987222 &Radio       \\        
16& CXB55 & 267.5735447 &-31.2430775 &12 &  [RDL94] Terzan 6 A          &267.574167 &-31.239722 &X           \\        
17& CXB58 & 268.5832235 &-29.6379212 &0.8  &  2MASS J17541996-2938157     &268.583188 &-29.637694 &EB*         \\        
18& CXB63 & 267.6738181 &-30.1941350 &1.3  &  Cl* NGC 6451     KF 227 &267.674208 &-30.194250 &in Cluster       \\       
19a& CXB93$^a$ & 266.5529013 &-32.1035349 &2.5  &  LTT  7072                   &266.552088 &-32.103529 &PM* M2~V        \\        
19b& CXB93$^a$ & 266.5529013 &-32.1035349 &3.7  &  ** LDS  611 / GJ  2130 C    &266.553167 &-32.102528 &**         \\        
20& CXB97 & 269.7613953 &-27.4890113 &0.9  &  V* V1723 Sgr                &269.761125 &-27.488917 &EB*WUMa     \\        
21& CXB100&  268.4645298& -29.650292 & 2.3 &   OGLE BUL-SC3 769186        & 268.464292& -29.650889& V*         \\         
22& CXB112&  263.2739071& -30.5863552& 2.0 &   LP  920-61                 & 263.274083& -30.585833& PM* M2.5      \\         
23& CXB116$^a$&  269.2814150& -27.1476849& 0.4 &   HD 314886                  & 269.281369& -27.147590& A5       \\         
24& CXB127&  266.9509625& -31.3875612& 3.0 &   NVSS J174748-312315        & 266.950958& -31.388389& Radio      \\         
25& CXB128$^a$&  266.7138646& -25.7794799& 1.5 &   CD-25 12283                & 266.714287& -25.779338& F8      \\         
26a& CXB150&  268.1381712& -29.9457729& 0.7 &   VCS4 J1752-2956            & 268.137946& -29.945806& Radio      \\         
26b& CXB150&  268.1381712& -29.9457729& 3.5 &   GCPS  115                  & 268.139292& -29.945750& Radio      \\         
27& CXB181$^a$&  268.73059000& -29.2027756& 0.3 &   HD 162962                  & 268.730569& -29.202854& A      \\         
28& CXB183&  268.6757225& -28.8307272& 3.0 &   IRAS 17515-2849            & 268.674792& -28.830500& Star       \\         
29& CXB200$^a$&  263.4644661& -30.8417862& 0.5 &   TYC 7376-433-1             & 263.464475& -30.841914& Star       \\         
30& CXB211$^a$&  265.8693744& -32.2325220& 2.6 &   HD 160826                  & 265.870188& -32.232264& B9~V       \\         
31& CXB225$^a$&  269.0803986& -28.4701699& 2.5 &   TYC 6853-3032-1            & 269.079825& -28.470642& Star       \\       
32& CXB233$^a$&  268.83897484& -28.5734201& 1.1 &   HD 316692                  & 268.839115& -28.573143& A0       \\         
33& CXB245&  268.2919765& -29.3556874& 0.5 &   OGLE J175310.04-292120.6   & 268.291833& -29.355722& Dwarf Nova  \\  
34a& CXB256&  267.7514663& -30.3199539& 1.7 &   Cl* NGC 6451    PMR 65& 267.751250& -30.319528& in Cluster      \\         
34b& CXB256&  267.7514663& -30.3199539& 1.7 &   Cl* NGC 6451    PMR 64& 267.751917& -30.319667& in Cluster      \\         
35& CXB287$^a$&  263.3901785& -30.534113 & 1.5 &   HD 158982                  & 263.389732& -30.533990& A2~IV/V      \\        
36& CXB293&  268.710370 & -29.3371961& 0.4 &   2MASS J17545048-2920142    & 268.710375& -29.337306& EB*        \\         
37& CXB302$^a$&  269.6706800& -27.9024008& 0.3 &   TYC 6849-1627-1            & 269.670621& -27.902478& Star       \\    
38& CXB306$^a$&  269.5399321& -28.1418302& 0.4 &   HD 163613                  & 269.539931& -28.141712& B1~Iab      \\         
39& CXB352&  268.4262642& -29.8320194& 1.3 &   OGLEII DIA BUL-SC3   5152  & 268.426333& -29.831667& EB*        \\         
40& CXB361&  268.1649063& -29.752345 & 5.0 &   OGLE BUL-SC37 441760       & 268.163375& -29.751944& V*         \\         
41& CXB366&  268.1003203& -29.7169994& 0.2 &   2MASS J17522407-2943013    & 268.100292& -29.717056& EB*        \\        
42& CXB380&  267.3212976& -31.2837757& 11.5&   SNR G358.4-01.9            & 267.325000& -31.283333& SNR        \\         
43a& CXB422$^a$&  262.8208422& -30.3215429& 0.8 &   HD 315956                  & 262.820644& -30.321404& F2      \\         
43b& CXB422$^a$&  262.8208422& -30.3215429& 3.2 &   [RHI84]  9-186             & 262.820125& -30.320917& M4       \\         
                                                                                                            
\end{tabular}
\end{center}
{\footnotesize $^a$ Association already found in \citet{2012ApJ...761..162H}  }                                                                                       
\end{table*}

Finally, we cross-correlated the positions of the CXB sources with the
entries in {\sc simbad} where we retained optical sources that have a
position within 5\arcsec\, of that of a CXB source and radio and X-ray
sources that have a position within 30\arcsec\, from a CXB source.
Table~\ref{simbad} contains the resulting list of sources. Some of the NVSS
sources are not found this way (cf.~with Table~\ref{radio}) whereas others
are (e.g.~the match between the NVSS source and CXB23 is also found using
{\sc simbad}). Many of the associations of CXB sources with bright optical
counterparts were already found in \citet{2012ApJ...761..162H}. Note that
some CXB sources have multiple entries as they have more than one potential
counterpart within 5\arcsec, such as CXB93, CXB256 and CXB422, or they have
multiple detections of presumably the same object with slightly different
positions such CXB9, CXB23 and CXB150.

In order to estimate the number of false positive identifications, we
then shifted all the CXB source positions by 15 or 30\arcsec\, north
or south, and we redid the cross-correlation.  On average, we get 5.5
{\sc simbad} matches, almost all of these spurious matches are OGLE sources,
with a few matches to stars from the open cluster NGC~6451.  So, we
therefore conclude that $\sim$38 of our 43 opt/IR matches are real
matches, with the OGLE matches being subject to the highest
false-alarm probability.

From Table~\ref{simbad}, we find three Cataclysmic Variables with
close positional matches to the CXB X-ray source positions (CXB10,
CXB26 and CXB245).  These associations are probably all real. Finally,
CXB97 is well matched with a W~UMa type source.  This is likely to be
a real match, and part of the predicted W~UMa population.

\section{Summary}

In this paper we have presented the \chan\, source list and some properties
of the X--ray sources of observations covering the $\approx$quarter of the
total survey area of 12 square degrees remaining to be done after the work of
\citet{2011ApJS..194...18J}. This paper thus completes the \chan\, survey
part of the Galactic Bulge Survey (GBS). The accurate \chan\, source position
will help identify the optical, near-infrared and UV counterparts. The 424
X--ray sources that have been discovered here, together with the 1216 unique
sources from \citet{2011ApJS..194...18J}, compares well with the total number
of $\approx 1650$ X--ray sources that we predicted we should detect in the
full 12 square degrees. However, this is of course no guarantee that the
number of sources per source class is close to the number we calculated.
Optical and near--infrared photometry including variability information and
spectroscopy is necessary to determine the nature of each of the sources (see
for instance \citealt{2013MNRAS.428.3543R}, \citealt{2013ApJ...769..120B}, 
\citealt{2013arXiv1310.2597H}, \citealt{2013arXiv1310.0224T}).

We discussed the apparent overdensity of sources in the (+l,-b)
quadrant of the GBS area. We conclude that this is caused by the lower
extinction in this quadrant.

We compared our source list with the source list of the RASS.  Furthermore,
we compared our \chan\, source list with the sources found in the catalog of
sources derived from pointed HRI and PSPC ROSAT observations that fall inside
the GBS area. Finally, we investigate whether some of the sources we report
on here are present in public radio surveys.

\section*{Acknowledgments} \noindent The authors would like to thank
the CXC for scheduling the \chan\, observations. R.I.H, C.T.B., C.B.J,
acknowledge support from the National Science Foundation under Grant
No. AST-0908789. This research has made use of the SIMBAD database,
operated at CDS, Strasbourg, France. This research has made use of the
VizieR catalogue access tool, CDS, Strasbourg, France. The original
description of the VizieR service was published in A\&AS 143, 23. This
publication makes use of data products from the Wide-field Infrared
Survey Explorer, which is a joint project of the University of
California, Los Angeles, and the Jet Propulsion Laboratory/California
Institute of Technology, funded by the National Aeronautics and Space
Administration (NASA). This work is based in part on observations made
with the Spitzer Space Telescope, which is operated by the Jet
Propulsion Laboratory, California Institute of Technology under a
contract with NASA.

\bibliographystyle{apj}

\begin{thebibliography}{}

\bibitem[\protect\citeauthoryear{{Arnaud}}{{Arnaud}}{1996}]{1996adass...5...17A}
{Arnaud}, K.~A. 1996, in ASP Conf. Ser. 101: Astronomical Data Analysis
  Software and Systems V, Vol.~5, 17

\bibitem[\protect\citeauthoryear{{Belczynski} \& {Taam}}{{Belczynski} \&
  {Taam}}{2004}]{2004ApJ...603..690B}
{Belczynski}, K.,  \& {Taam}, R.~E. 2004, \apj, 603, 690

\bibitem[\protect\citeauthoryear{{Belczynski} et~al.}{{Belczynski}
  et~al.}{2012}]{2012ApJ...757...91B}
{Belczynski}, K., {Wiktorowicz}, G., {Fryer}, C.~L., {Holz}, D.~E.,  \&
  {Kalogera}, V. 2012, \apj, 757, 91

\bibitem[\protect\citeauthoryear{{Benjamin} et~al.}{{Benjamin}
  et~al.}{2003}]{2003PASP..115..953B}
{Benjamin}, R.~A., et~al. 2003, \pasp, 115, 953

\bibitem[\protect\citeauthoryear{{Britt} et~al.}{{Britt}
  et~al.}{2013}]{2013ApJ...769..120B}
{Britt}, C.~T., et~al. 2013, \apj, 769, 120

\bibitem[\protect\citeauthoryear{{Castelli} \& {Kurucz}}{{Castelli} \&
  {Kurucz}}{2004}]{2004astro.ph..5087C}
{Castelli}, F.,  \& {Kurucz}, R.~L. 2004, ArXiv Astrophysics e-prints

\bibitem[\protect\citeauthoryear{{Churchwell} et~al.}{{Churchwell}
  et~al.}{2009}]{2009PASP..121..213C}
{Churchwell}, E., et~al. 2009, \pasp, 121, 213

\bibitem[\protect\citeauthoryear{{Evans} et~al.}{{Evans}
  et~al.}{2010}]{2010ApJS..189...37E}
{Evans}, I.~N., et~al. 2010, \apjs, 189, 37

\bibitem[\protect\citeauthoryear{{Farr} et~al.}{{Farr}
  et~al.}{2011}]{2011ApJ...741..103F}
{Farr}, W.~M., {Sravan}, N., {Cantrell}, A., {Kreidberg}, L., {Bailyn}, C.~D.,
  {Mandel}, I.,  \& {Kalogera}, V. 2011, \apj, 741, 103

\bibitem[\protect\citeauthoryear{{Fryer} et~al.}{{Fryer}
  et~al.}{2012}]{2012ApJ...749...91F}
{Fryer}, C.~L., {Belczynski}, K., {Wiktorowicz}, G., {Dominik}, M., {Kalogera},
  V.,  \& {Holz}, D.~E. 2012, \apj, 749, 91

\bibitem[\protect\citeauthoryear{{Garmire}}{{Garmire}}{1997}]{1997AAS...190.3404G}
{Garmire}, G.~P. 1997, in Bulletin of the American Astronomical Society,
  Vol.~29, American Astronomical Society Meeting Abstracts \#190, 823

\bibitem[\protect\citeauthoryear{{Gehrels}}{{Gehrels}}{1986}]{1986ApJ...303..336G}
{Gehrels}, N. 1986, \apj, 303, 336

\bibitem[\protect\citeauthoryear{{Gonzalez} et~al.}{{Gonzalez}
  et~al.}{2012}]{2012A&A...543A..13G}
{Gonzalez}, O.~A., {Rejkuba}, M., {Zoccali}, M., {Valenti}, E., {Minniti}, D.,
  {Schultheis}, M., {Tobar}, R.,  \& {Chen}, B. 2012, \aap, 543, A13

\bibitem[\protect\citeauthoryear{{Grimm} et~al.}{{Grimm}
  et~al.}{2005}]{2005ApJS..161..271G}
{Grimm}, H., {McDowell}, J., {Zezas}, A., {Kim}, D.,  \& {Fabbiano}, G. 2005,
  \apjs, 161, 271

\bibitem[\protect\citeauthoryear{{Hong} et~al.}{{Hong}
  et~al.}{2009}]{2009ApJ...706..223H}
{Hong}, J.~S., {van den Berg}, M., {Grindlay}, J.~E.,  \& {Laycock}, S. 2009,
  \apj, 706, 223

\bibitem[\protect\citeauthoryear{{Hynes} et~al.}{{Hynes}
  et~al.}{2013}]{2013arXiv1310.2597H}
{Hynes}, R.~I., et~al. 2013, ArXiv e-prints

\bibitem[\protect\citeauthoryear{{Hynes} et~al.}{{Hynes}
  et~al.}{2012}]{2012ApJ...761..162H}
{Hynes}, R.~I., et~al. 2012, \apj, 761, 162

\bibitem[\protect\citeauthoryear{{Ivanova} et~al.}{{Ivanova}
  et~al.}{2013}]{2013A&ARv..21...59I}
{Ivanova}, N., et~al. 2013, \aapr, 21, 59

\bibitem[\protect\citeauthoryear{{Jonker} et~al.}{{Jonker}
  et~al.}{2011}]{2011ApJS..194...18J}
{Jonker}, P.~G., et~al. 2011, \apjs, 194, 18

\bibitem[\protect\citeauthoryear{{Kiel} \& {Hurley}}{{Kiel} \&
  {Hurley}}{2006}]{2006MNRAS.369.1152K}
{Kiel}, P.~D.,  \& {Hurley}, J.~R. 2006, \mnras, 369, 1152

\bibitem[\protect\citeauthoryear{{Kim} et~al.}{{Kim}
  et~al.}{2004}]{2004ApJS..150...19K}
{Kim}, D., et~al. 2004, \apjs, 150, 19

\bibitem[\protect\citeauthoryear{{King} \& {Ritter}}{{King} \&
  {Ritter}}{1999}]{1999MNRAS.309..253K}
{King}, A.~R.,  \& {Ritter}, H. 1999, \mnras, 309, 253

\bibitem[\protect\citeauthoryear{{Kreidberg} et~al.}{{Kreidberg}
  et~al.}{2012}]{2012ApJ...757...36K}
{Kreidberg}, L., {Bailyn}, C.~D., {Farr}, W.~M.,  \& {Kalogera}, V. 2012, \apj,
  757, 36

\bibitem[\protect\citeauthoryear{{Lasota}}{{Lasota}}{2008}]{2008NewAR..51..752L}
{Lasota}, J.-P. 2008, New Astronomy Review, 51, 752

\bibitem[\protect\citeauthoryear{{Lee}, {Brown}, \& {Wijers}}{{Lee}
  et~al.}{2002}]{2002ApJ...575..996L}
{Lee}, C.-H., {Brown}, G.~E.,  \& {Wijers}, R.~A.~M.~J. 2002, \apj, 575, 996

\bibitem[\protect\citeauthoryear{{Maccarone} et~al.}{{Maccarone}
  et~al.}{2012}]{2012MNRAS.426.3057M}
{Maccarone}, T.~J., et~al. 2012, \mnras, 426, 3057

\bibitem[\protect\citeauthoryear{{Murray} et~al.}{{Murray}
  et~al.}{2005}]{2005ApJS..161....1M}
{Murray}, S.~S., et~al. 2005, \apjs, 161, 1

\bibitem[\protect\citeauthoryear{{Narayan} \& {McClintock}}{{Narayan} \&
  {McClintock}}{2005}]{2005ApJ...623.1017N}
{Narayan}, R.,  \& {McClintock}, J.~E. 2005, \apj, 623, 1017

\bibitem[\protect\citeauthoryear{{Nord} et~al.}{{Nord}
  et~al.}{2004}]{2004AJ....128.1646N}
{Nord}, M.~E., {Lazio}, T.~J.~W., {Kassim}, N.~E., {Hyman}, S.~D., {LaRosa},
  T.~N., {Brogan}, C.~L.,  \& {Duric}, N. 2004, \aj, 128, 1646

\bibitem[\protect\citeauthoryear{{{\"O}zel} et~al.}{{{\"O}zel}
  et~al.}{2010}]{2010ApJ...725.1918O}
{{\"O}zel}, F., {Psaltis}, D., {Narayan}, R.,  \& {McClintock}, J.~E. 2010,
  \apj, 725, 1918

\bibitem[\protect\citeauthoryear{{{\"O}zel} et~al.}{{{\"O}zel}
  et~al.}{2012}]{2012ApJ...757...55O}
{{\"O}zel}, F., {Psaltis}, D., {Narayan}, R.,  \& {Santos Villarreal}, A. 2012,
  \apj, 757, 55

\bibitem[\protect\citeauthoryear{{Pfahl}, {Rappaport}, \&
  {Podsiadlowski}}{{Pfahl} et~al.}{2002}]{2002ApJ...571L..37P}
{Pfahl}, E., {Rappaport}, S.,  \& {Podsiadlowski}, P. 2002, \apjl, 571, L37

\bibitem[\protect\citeauthoryear{{Predehl} \& {Schmitt}}{{Predehl} \&
  {Schmitt}}{1995}]{1995A&A...293..889P}
{Predehl}, P.,  \& {Schmitt}, J.~H.~M.~M. 1995, \aap, 293, 889

\bibitem[\protect\citeauthoryear{{Rappaport} et~al.}{{Rappaport}
  et~al.}{1994}]{1994ApJ...423..633R}
{Rappaport}, S., {Dewey}, D., {Levine}, A.,  \& {Macri}, L. 1994, \apj, 423,
  633

\bibitem[\protect\citeauthoryear{{Ratti} et~al.}{{Ratti}
  et~al.}{2013}]{2013MNRAS.428.3543R}
{Ratti}, E.~M., et~al. 2013, \mnras, 428, 3543

\bibitem[\protect\citeauthoryear{{Schlegel}, {Finkbeiner}, \&
  {Davis}}{{Schlegel} et~al.}{1998}]{1998ApJ...500..525S}
{Schlegel}, D.~J., {Finkbeiner}, D.~P.,  \& {Davis}, M. 1998, \apj, 500, 525

\bibitem[\protect\citeauthoryear{{Shahbaz} \& {Kuulkers}}{{Shahbaz} \&
  {Kuulkers}}{1998}]{1998MNRAS.295L...1S}
{Shahbaz}, T.,  \& {Kuulkers}, E. 1998, \mnras, 295, L1

\bibitem[\protect\citeauthoryear{{Skrutskie} et~al.}{{Skrutskie}
  et~al.}{2006}]{2006AJ....131.1163S}
{Skrutskie}, M.~F., et~al. 2006, \aj, 131, 1163

\bibitem[\protect\citeauthoryear{{Snowden} et~al.}{{Snowden}
  et~al.}{1997}]{1997ApJ...485..125S}
{Snowden}, S.~L., et~al. 1997, \apj, 485, 125

\bibitem[\protect\citeauthoryear{{Torres} et~al.}{{Torres}
  et~al.}{2013}]{2013arXiv1310.0224T}
{Torres}, M.~A.~P., et~al. 2013, ArXiv e-prints

\bibitem[\protect\citeauthoryear{{Udalski} et~al.}{{Udalski}
  et~al.}{2012}]{2012AcA....62..133U}
{Udalski}, A., et~al. 2012, Acta Astronomica \& arXiv 1208.0369, 62, 133

\bibitem[\protect\citeauthoryear{{Voges} et~al.}{{Voges}
  et~al.}{1999}]{1999A&A...349..389V}
{Voges}, W., et~al. 1999, \aap, 349, 389

\bibitem[\protect\citeauthoryear{{Wang}, {Gotthelf}, \& {Lang}}{{Wang}
  et~al.}{2002}]{2002Natur.415..148W}
{Wang}, Q.~D., {Gotthelf}, E.~V.,  \& {Lang}, C.~C. 2002, \nat, 415, 148

\bibitem[\protect\citeauthoryear{{Warwick}, {P{\'e}rez-Ram{\'{\i}}rez}, \&
  {Byckling}}{{Warwick} et~al.}{2011}]{2011MNRAS.413..595W}
{Warwick}, R.~S., {P{\'e}rez-Ram{\'{\i}}rez}, D.,  \& {Byckling}, K. 2011,
  \mnras, 413, 595

\bibitem[\protect\citeauthoryear{{Weisskopf} et~al.}{{Weisskopf}
  et~al.}{2002}]{2002PASP..114....1W}
{Weisskopf}, M.~C., {Brinkman}, B., {Canizares}, C., {Garmire}, G., {Murray},
  S.,  \& {Van Speybroeck}, L.~P. 2002, \pasp, 114, 1

\bibitem[\protect\citeauthoryear{{Wright} et~al.}{{Wright}
  et~al.}{2010}]{2010AJ....140.1868W}
{Wright}, E.~L., et~al. 2010, \aj, 140, 1868

\bibitem[\protect\citeauthoryear{{Wu} et~al.}{{Wu}
  et~al.}{2010}]{2010ApJ...718..620W}
{Wu}, Y.~X., {Yu}, W., {Li}, T.~P., {Maccarone}, T.~J.,  \& {Li}, X.~D. 2010,
  \apj, 718, 620

\end{thebibliography}

\end{document}